\documentclass[12pt]{article}
\usepackage{graphicx}

\begin{document}

\baselineskip=18pt plus 1pt minus 1pt
\begin{center} 

{\large\bf Parameter-free solution of the Bohr Hamiltonian for actinides
critical in the octupole mode} 
\bigskip

{D. Lenis\footnote{e-mail: lenis@inp.demokritos.gr} and 
Dennis Bonatsos\footnote{e-mail: bonat@inp.demokritos.gr}}  
\bigskip

{Institute of Nuclear Physics, N.C.S.R. ``Demokritos''}

{GR-15310 Aghia Paraskevi, Attiki, Greece}

\bigskip

{\bf Abstract}

\end{center} 

An analytic, parameter-free (up to overall scale factors) solution of the 
Bohr Hamiltonian involving axially symmetric quadrupole and octupole 
deformations, as well as an infinite well potential, is obtained, after 
separating variables in a way reminiscent of the Variable Moment of Inertia
(VMI) concept. Normalized spectra and $B(EL)$ ratios are found to agree 
with experimental data for $^{226}$Ra and $^{226}$Th, the nuclei known to lie 
closest to the border between octupole deformation and octupole vibrations 
in the light actinide region.

\bigskip\bigskip

{\bf 1. Introduction}

Critical point symmetries \cite{IacE5,IacX5} are attracting recently 
considerable interest, since they provide parameter-independent (up to 
overall scale factors) predictions supported by experiment 
\cite{CZE5,ClarkE5,CZX5,ClarkX5}. The E(5) \cite{IacE5} and X(5) \cite{IacX5} 
critical point symmetries have been obtained from the Bohr Hamiltonian 
\cite{Bohr} after separating variables in different ways and using an 
infinite square well potential in the $\beta$ (quadrupole)
variable, the latter corresponding to the critical point of the transition 
from quadrupole vibrations [{\rm U(5)}] to axial quadrupole deformation 
[{\rm SU(3)}] \cite{IacX5}. A systematic study of phase transitions in 
nuclear collective models has been given in \cite{RoweI,RoweII,RoweIII}.

In the present work a solution of the Bohr Hamiltonian aiming at the 
description of the transition from axial octupole deformation to octupole 
vibrations in the light actinides \cite{BN} is worked out. In the spirit of 
E(5) and X(5) the solution involves an infinite square well potential in the 
deformation variable and leads to parameter-free (up to overal scale factors) 
predictions for spectra and $B(EL)$ transition rates. Both (axially symmetric)
quadrupole 
and octupole deformations are taken into account, in order to describe
low-lying negative parity states related to octupole deformation, known 
to occur in the light actinides \cite{BN}. Separation of variables is 
achieved in a novel way, reminiscent of the Variable Moment of Inertia 
(VMI) concept \cite{VMI}. The parameter-free predictions of the model turn out
to be in good agreement with experimental data for $^{226}$Ra \cite{Cocks}
and $^{226}$Th \cite{NDS}, 
the nuclei known \cite{Bizzeti,AQOA}
to lie closest to the transition from octupole deformation
to octupole vibrations in this region.  

In Section 2 of the present work the solution is worked out. Numerical 
results are given and compared to experimental data in Section 3, while 
Section 4 contains discussion of the present results and plans for further 
work.  

{\bf 2. The model} 

When only axially symmetric quadrupole ($\beta_2$) and octupole ($\beta_3$) 
deformations
are taken into account, the Hamiltonian reads \cite{Dzy,Den}
\begin{equation}\label{eq:e1}
H = -\sum_{\lambda=2,3} {\hbar^2 \over 2 B_\lambda} {1\over \beta_\lambda^3} 
{\partial \over \partial \beta_\lambda} \beta_\lambda^3 {\partial \over 
\partial \beta_\lambda} + {\hbar^2 \hat {L^2} \over 6(B_2 \beta_2^2 + 
2 B_3 \beta_3^2) } + V(\beta_2,\beta_3)  
\end{equation}
where $B_2$, $B_3$ are the mass parameters. 

One then seeks solutions of the Schr\"odinger equation of the form \cite{Dzy}
\begin{equation}\label{eq:e2}
\Phi^{\pm}_L(\beta_2,\beta_3,\theta) = (\beta_2 \beta_3)^{-3/2} 
\Psi^{\pm}_L(\beta_2,\beta_3) \vert LM0,\pm\rangle,
\end{equation} 
where $M$ is the angular momentum projection onto the 
laboratory-fixed $\hat z$-axis, $K=0$ is the projection onto the body-fixed 
$\hat z'$-axis, and the functions $\vert LM0, +\rangle$ and 
$\vert LM0,-\rangle$ transform according to the irreducible 
representations (irreps) {\rm A} and {\rm B}$_1$ of the group {\rm D}$_2$
respectively \cite{Dzy,Den}, their explicit form being given 
in \cite{AQOA,BM}.

Introducing \cite{Dzy,Den} 
\begin{equation}\label{eq:e3}
\tilde \beta_2 = \beta_2 \sqrt{B_2\over B}, \quad \tilde \beta_3 = \beta_3 
\sqrt{B_3\over B}, \quad B= {B_2+B_3 \over 2}, 
\end{equation}
reduced energies $\epsilon=(2B/\hbar^2) E$ and reduced potentials 
$u=(2B/\hbar^2) V$ \cite{IacE5,IacX5}, as well as polar coordinates 
(with $0\leq \tilde \beta < \infty$ and $-\pi/2 \leq \phi \leq \pi/2$) 
\cite{Dzy,Den} 
\begin{equation}\label{eq:e4} 
\tilde \beta_2 = \tilde \beta \cos \phi, \quad \tilde \beta_3 = \tilde \beta
\sin \phi, \quad \tilde \beta = \sqrt{\tilde \beta_2^2 + \tilde \beta_3^2}, 
\end{equation} 
the Schr\"odinger equation takes the form \cite{AQOA} 
\begin{equation}\label{eq:e5}
\left[ -{\partial^2 \over \partial \tilde \beta^2} -{1\over \tilde \beta} 
{\partial \over \partial \tilde \beta} +{L(L+1) \over 3 \tilde \beta^2 
(1+\sin^2\phi) } -{1\over \tilde \beta^2} {\partial^2 \over \partial \phi^2} 
+ u(\tilde \beta,\phi) + {3\over  \tilde \beta^2 \sin^2 2\phi}-\epsilon_L
\right] \Psi_L^{\pm}(\tilde \beta,\phi) =0. 
\end{equation}

Separation of variables in Eq. (\ref{eq:e5}) can be achieved by assuming 
the potential to be of the form \cite{Dzy,Fortun}
$u(\tilde \beta,\phi) = u(\tilde\beta) + u(\phi)/\tilde \beta^2$,
leading to 
\begin{equation}\label{eq:e6}
\tilde \beta^2 \left( -{\partial^2 \over \partial \tilde \beta^2} -
{1\over \tilde \beta} {\partial \over \partial \tilde \beta} +u(\tilde \beta)
-\epsilon_{\tilde \beta}(L) \right) \psi^\pm_L(\tilde \beta) =
-\nu^2 \psi^\pm_L(\tilde \beta), 
\end{equation}
\begin{equation}\label{eq:e7}
\left( {\partial^2 \over \partial \phi^2} -u(\phi)-u_L(\phi)\right)
\chi^\pm (\phi) = -\nu^2 \chi^\pm(\phi), 
\end{equation}
where 
\begin{equation}\label{eq:e8}
u_L(\phi) = {3\over \sin^2 2\phi} + {L(L+1) \over 3(1+\sin^2 \phi)}, 
\end{equation}
with $\nu^2$ being the separation constant and 
$\Psi^{\pm}_L(\tilde \beta,\phi) = \psi^{\pm}_L(\tilde \beta) \chi^{\pm} 
(\phi)$, where, however, the $\pm$ indices have become redundant.   

The potential $u_L(\phi)$ of Eq. (\ref{eq:e8}) is shown in Fig. 1 for 
several values of $L$, normalized to its minimum value for each $L$. 
It is clear that in each case the potential has the form of a deep well, 
possessing an $L$-dependent minimum, denoted by $\phi_L$ and determined from 
the equation
\begin{equation}\label{eq:e9}
u'_L(\phi) = -{2\over 3} {L(L+1) \sin\phi \cos\phi \over (1+\sin^2\phi)^2}
-{12 \cos 2\phi\over \sin^3 2\phi} =0. 
\end{equation}  
Using standard trigonometric identities and defining $x=\sin^2\phi$ and 
$b=9/(4L(L+1))$ one easily sees that Eq. (\ref{eq:e9}) takes the form 
\begin{equation}\label{eq:e9a}
x^4-2(1+b)x^3 +(1-3b)x^2+b=0,  
\end{equation}
which turns out to have only one real root in the interval $0\leq x\leq1$
(imposed by $x=\sin^2\phi$).

Given its form, the potential $u_L(\phi)$ can be approximated around 
the minimum by the first terms of the Taylor expansion as 
\begin{equation}\label{eq:e10} 
u_L(\phi) \approx u_L(\phi_L) + {u''_L(\phi_L)\over 2} (\phi-\phi_L)^2.
\end{equation}
In Eq. (\ref{eq:e7}) one can then omit the potential $u(\phi)$, treating 
$u_L(\phi)$ as an effective potential naturally occuring in the framework 
of the theory, leading Eq. (\ref{eq:e7}) into the harmonic oscillator form
\begin{equation}\label{eq:e11}
-{\partial^2 \chi \over \partial \zeta^2} +\zeta^2 \chi = \varepsilon_L \chi, 
\end{equation} 
with 
\begin{equation}\label{eq:e12} 
\zeta^2 = \sqrt{ {u''_L(\phi_L)\over 2} } (\phi-\phi_L)^2 , \qquad 
\varepsilon_L= {\nu^2-u_L(\phi_L) \over \sqrt{ { u''_L(\phi_L) \over 2}}} .
\end{equation}
Since $\varepsilon_L=2n+1$, where $n$ is the number of oscillator quanta,
one obtains 
\begin{equation}\label{eq:e13}
\nu^2 = \sqrt{u''_L(\phi_L) \over 2} (2n+1) + u_L(\phi_L). 
\end{equation} 
In what follows we are going to be limited to the case $n=0$. 

Returning to Eq. (\ref{eq:e6}), using for $u(\tilde \beta)$ an infinite well 
potential ($u(\tilde\beta)= 0$ if $\tilde\beta \leq \tilde \beta_W$; 
 $u(\tilde \beta)=\infty$ if $\tilde \beta > \tilde \beta_W$),
and defining \cite{IacX5} $\epsilon_{\tilde \beta} = k_{\tilde\beta}^2$,  
$z=\tilde \beta k_{\tilde \beta}$, one is led to the Bessel equation 
\begin{equation}\label{eq:e14} 
{d^2\psi_\nu \over dz^2} + {1\over z} {d\psi_\nu \over dz} 
+\left [ 1- {\nu^2 \over z^2} \right] \psi_\nu =0. 
\end{equation}
Then the boundary condition $\psi_\nu(\tilde \beta_W)=0$ determines 
the spectrum 
\begin{equation}\label{eq:e15} 
\epsilon_{\tilde \beta, s, \nu} = \epsilon_{\tilde \beta, s, L} 
= (k_{s,\nu})^2, \qquad k_{s,\nu}= {x_{s,\nu} \over \tilde \beta_W}, 
\end{equation}
and the eigenfunctions 
\begin{equation}\label{eq:e16} 
\psi_{s,\nu}(\tilde\beta) = \psi_{s,L}(\tilde \beta) = 
c_{s,\nu} J_{\nu}(k_{s,\nu} \tilde \beta), 
\end{equation}
where $x_{s,\nu}$ is the $s$th zero of the Bessel function $J_\nu(z)$, 
while $c_{s,\nu}$ are normalization constants, determined from the 
condition $\int_0^{\tilde \beta_W} 
\vert \psi_{s,\nu}(\tilde \beta)\vert^2 \tilde \beta 
d\tilde \beta =1$ to be $c_{s,\nu}=\sqrt{2}/J_{\nu+1}(k_{s,\nu})$. 
The notation has been kept similar to that of Ref. \cite{IacX5}. 

A few comments are now in place:

a) $L$-dependent potentials, as the one of Eq. (\ref{eq:e8}), 
are known to occur in nuclear physics in the framework of the optical 
model potential \cite{Fiedel}, as well as in the study of quasimolecular 
resonances, such as $^{12}$C+$^{12}$C \cite{Scheid}.

b) The procedure followed for the determination of $\phi_L$ is reminiscent 
of the Variable Moment of Inertia (VMI) model \cite{VMI}. In the VMI case 
the energy is minimized with respect to the moment of inertia for each 
$L$ separately, resulting in a moment of inertia increasing with $L$. 
In the present case the effective potential energy $u_L$ is minimized with 
respect 
to $\phi$ for each $L$ separately, resulting in $\phi_L$ values increasing 
with $L$. As a consequence, in Eq. (\ref{eq:e8}) the denominator of $L(L+1)$, 
which can be considered roughly as playing the role of a moment of inertia, 
is also increasing with $L$.  

c) In order to separate variables, one can in general assume 
$u(\beta,\gamma) = u(\beta)+u(\gamma)$, as in the X(5) model \cite{IacX5}, 
or $u(\beta,\gamma)=u(\beta)+u(\gamma)/\beta^2$, as in Refs. \cite{Dzy,Fortun}.
In the former case the separation is approximate, 
since a $\beta^2$ term is involved in the $\gamma$ equation, replaced 
by its average value, but no extra parameter is introduced in the 
$\beta$-equation by the separation.
In the latter case, the separation is exact, but (at least) 
one extra parameter appears in the $\beta$-equation, coming in from 
the $\gamma$-equation through the separation constant. In the present 
model this disadvantage of the latter case is avoided through the VMI-like 
procedure adapted.    

d) For each $L$ the specific value of the variable $\phi$, which decides 
the relative presence of the quadrupole and octupole deformations,
is determined in Eq. (\ref{eq:e7}) by the effective potential $u_L(\phi)$,
which has a rigid shape, as seen in Fig. 1, while the potential $u(\phi)$
plays no role.  
 
The calculation of $B(EL)$ transition rates proceeds as in Ref. \cite{AQOA} 
and need not be repeated here.  

{\bf 3. Numerical results and comparison to experiment} 

The parameter-free predictions of the model for the lowest bands are given 
in Table 1, together with the experimental spectra of $^{226}$Ra \cite{Cocks}
and $^{226}$Th \cite{NDS}, 
which are known to lie near the border between the regions of octupole 
deformation and octupole vibrations \cite{Bizzeti,AQOA}, as also seen 
in Figs. 2(a) and 2(b), where the experimental spectra of $^{218-228}$Ra 
and $^{220-234}$Th are included. In both figures the region below the 
theoretical predictions corresponds to octupole deformation, characterized 
by minimal odd-even staggering rapidly decreasing and disappearing with 
increasing $L$, while the region above the theoretical 
predictions corresponds to octupole vibrations, characterized by large 
odd-even staggering decreasing very slowly with increasing $L$. 
As seen in Table 1, in the case of $^{226}$Ra and $^{226}$Th the odd-even 
staggering is non-negligible only for the lowest four odd levels, the 
agreement between theory and experiment being very good for the even 
levels, as well as with the rest of the odd ones. The absence of staggering 
in the present model is due to the fact that the (infinite) potential wells 
for $\beta_3<0$ and $\beta_3>0$ [see Eq. (\ref{eq:e4})]
are separated by an infinite barrier, and not by a finite one, as needed 
for odd-even staggering to be present \cite{Jolos}. 

Several parameter-free predictions for $B(E1)$, $B(E2)$, and $B(E3)$ 
transition rates, appropriately normalized, are reported in Table 2. 
Since the lack of experimental data does not allow for direct comparison 
to experiment, comparisons to $B(E1)$ branching ratios of $^{226}$Ra
\cite{Woll} and $B(E1)/B(E2)$ ratios for $^{226}$Th \cite{Bizzeti} are shown 
in Figs. 3(a) and 3(b)
respectively. The theoretical predictions in Fig. 3(a) are compatible with 
the data within the experimental errors, while in addition they are quite 
similar to predictions of the Extended Coherent States Model (ECSM) 
\cite{Raduta} obtained with the lowest order $E1$ operator (R-h), 
as well as with two different choices of the $E1$ operator including 
anharmonicities (R-I, R-II) \cite{Raduta2}. The theoretical predictions 
in Fig. 3(b) 
are also compatible with the data within the experimental errors, lying 
considerably lower than the predictions of Ref. \cite{Bizzeti} (BBS).   

It is worth remarking that the ground state band spectrum and intraband 
$B(E2)$s of the present model are quite similar to these of the X(5) model
\cite{IacX5} (extensively tabulated in Ref. \cite{X5}). 
Indeed the present model can be considered as an extension 
of X(5), in which the octupole 
degree of freedom is taken into account in order to account for the 
low-lying negative parity bands, while in parallel the $\gamma$ 
degree of freedom is left out in order to keep the problem tractable.
One important difference between the two models is related to the (normalized)
position of the $0_2^+$ state, which is predicted at 5.649 by the X(5) 
model \cite{IacX5}, but at 12.569 by the present model. This implies that 
while searching for X(5)-like nuclei in the light actinide region, one should 
expect the $0_2^+$ state to occur higher by a factor of two. 
Indeed the Ra and Th isotopes near $A=226$ exhibit high-lying $0_2^+$ 
states \cite{Cocks,NDS}.   

It should also be noticed that the predictions of the present 
parameter-independent model are very similar to those of the one-parameter
($\phi_0$) Analytic Quadrupole Octupole Axially symmetric (AQOA) model 
\cite{AQOA}, in which best agreement to experiment is obtained for 
$\phi_0=56^{\rm o}$ in the case of $^{226}$Ra, and for $\phi_0=60^{\rm o}$
in the case of $^{226}$Th. These values of $\phi_0$ are understandable, 
when compared to the $\phi_L$ values shown in Table 1. 
 
{\bf 4. Discussion}

A parameter-free (up to overall scale factors) version of an Analytic 
Quadrupole Octupole Axially symmetric (AQOA) model involving an infinite well
potential, suitable for describing 
the transition from octupole deformation to octupole vibrations in the 
light actinides, has been constructed, 
after separating variables in the Bohr Hamiltonian in a way reminiscent of the
Variable Moment of Inertia concept. Spectra and $B(EL)$ ratios are shown 
to be in good agreement with experimental data for $^{226}$Ra and $^{226}$Th, 
the nuclei supposed to lie closest to the above mentioned border. 
Application of the model in the $A\approx 150$ region, where octupole 
deformation is also well established \cite{BN}, is receiving attention.

\parindent=0pt

\begin{table}

\caption{Spectra of the present model for the ground state band and the 
associated negative 
parity band ($s=1$), as well as for the first excited band ($s=2$),
together with relevant values of $\phi_L$ and experimental data 
for $^{226}$Ra \cite{Cocks} and $^{226}$Th \cite{NDS}.
Each spectrum is normalized to the energy of its own $2_1^+$ state.
See section 3 for further discussion.} 

\bigskip

\begin{tabular}{r r r r r r r r r r }
\hline
$L^{\pi}$ & $\phi_L$ & th & $^{226}$Ra & $^{226}$Th & 
$L^{\pi}$ & $\phi_L$ & th & $^{226}$Ra & $^{226}$Th \\
 \hline
$s=1$ &    &       &       &       &       &   &   &   &  \\
$0^+$  & 45.00& 0.000 & 0.000 & 0.000 & $1^-$ & 45.70& 0.337 & 3.747 & 3.191 \\
$2^+$  & 47.01& 1.000 & 1.000 & 1.000 & $3^-$ & 48.77& 1.967 & 4.749 & 4.259 \\
$4^+$  & 50.77& 3.200 & 3.127 & 3.136 & $5^-$ & 52.79& 4.657 & 6.603 & 6.240 \\
$6^+$  & 54.71& 6.297 & 6.155 & 6.195 & $7^-$ & 56.47& 8.093 & 9.264 & 9.112 \\
$8^+$  & 58.05&10.025 & 9.891 & 9.999 & $9^-$ & 59.46&12.081 &12.677 &12.785 \\
$10^+$ & 60.73&14.254 &14.185 &14.409 &$11^-$ & 61.86&16.537 &16.743 &17.152 \\
$12^+$ & 62.89&18.928 &18.931 &19.324 &$13^-$ & 63.81&21.424 &21.388 &22.105 \\
$14^+$ & 64.65&24.023 &24.061 &24.675 &$15^-$ & 65.42&26.724 &26.536 &27.554 \\
$16^+$ & 66.13&29.526 &29.523 &30.413 &$17^-$ & 66.78&32.427 &32.126 &33.418 \\
$18^+$ & 67.38&35.428 &35.300 &36.497 &$19^-$ & 67.94&38.527 &38.099 &39.627 \\
$20^+$ & 68.46&41.724 &41.375 &42.896 &       &      &       &       &       \\
\hline
$s=2$ &    &       &       &       &       &    &   \\
$0^+$  & 45.00&12.569 &12.186 &11.152 &       &      &       &       &       \\
$2^+$  & 47.01&14.253 &       &       &       &      &       &       &       \\
$4^+$  & 50.77&17.871 &       &       &       &      &       &       &       \\
\hline
\end{tabular}
\end{table}

\parindent=0pt

\begin{table}

\caption{ $B(EL;L_i \to L_f)$ values between states of the present model with 
$s=1$. 
$B(E2)$s with $L_i$ and $L_f$ even are normalized to the $2_1^+\to 0_1^+$ 
transition, while $B(E2)$s with $L_i$ and $L_f$ odd are normalized to the 
$3_1^-\to 1_1^-$ transition. $B(E1)$s are normalized to the $1_1^- \to 0_1^+$
transition, while $B(E3)$s are normalized to the $3_1^-\to 0_1^+$ transition.  
See section 3 for further discussion.} 

\bigskip

\begin{tabular}{r r r r r r r r r r r r}
\hline
$L_i^{\pi}$ & $L_f^{\pi}$ &  $B(E2)$ &
$L_i^{\pi}$ & $L_f^{\pi}$ &  $B(E1)$ &
$L_i^{\pi}$ & $L_f^{\pi}$ &  $B(E3)$ 
$L_i^{\pi}$ & $L_f^{\pi}$ &  $B(E3)$ \\
 \hline
$ 2^+$ &$ 0^+$ & 1.000 &$ 1^-$ &$ 0^+$ & 1.000 &$ 3^-$ &$ 0^+$ & 1.000 &
$ 2^+$ &$ 1^-$ & 1.793 \\
$ 4^+$ &$ 2^+$ & 1.494 &$ 2^+$ &$ 1^-$ & 1.238 &$ 4^+$ &$ 1^-$ & 1.358 &
$ 3^-$ &$ 2^+$ & 1.362 \\
$ 6^+$ &$ 4^+$ & 1.749 &$ 3^-$ &$ 2^+$ & 1.389 &$ 5^-$ &$ 2^+$ & 1.581 &
$ 4^+$ &$ 3^-$ & 1.341 \\
$ 8^+$ &$ 6^+$ & 1.944 &$ 4^+$ &$ 3^-$ & 1.522 &$ 6^+$ &$ 3^-$ & 1.752 &
$ 5^-$ &$ 4^+$ & 1.369 \\
$10^+$ &$ 8^+$ & 2.103 &$ 5^-$ &$ 4^+$ & 1.650 &$ 7^-$ &$ 4^+$ & 1.896 &
$ 6^+$ &$ 5^-$ & 1.411 \\
$12^+$ &$10^+$ & 2.237 &$ 6^+$ &$ 5^-$ & 1.775 &$ 8^+$ &$ 5^-$ & 2.022 &
$ 7^-$ &$ 6^+$ & 1.456 \\
$14^+$ &$12^+$ & 2.350 &$ 7^-$ &$ 6^+$ & 1.896 &$ 9^-$ &$ 6^+$ & 2.136 &
$ 8^+$ &$ 7^-$ & 1.500 \\
$16^+$ &$14^+$ & 2.448 &$ 8^+$ &$ 7^-$ & 2.012 &$10^+$ &$ 7^-$ & 2.238 &
$ 9^-$ &$ 8^+$ & 1.542 \\
$18^+$ &$16^+$ & 2.534 &$ 9^-$ &$ 8^+$ & 2.123 &$11^-$ &$ 8^+$ & 2.331 &
$10^+$ &$ 9^-$ & 1.582 \\
$20^+$ &$18^+$ & 2.611 &$10^+$ &$ 9^-$ & 2.228 &$12^+$ &$ 9^-$ & 2.416 &
$11^-$ &$10^+$ & 1.620 \\
       &       &       &$11^-$ &$10^+$ & 2.328 &$13^-$ &$10^+$ & 2.494 &
$12^+$ &$11^-$ & 1.655 \\
$ 3^-$ &$ 1^-$ & 1.000 &$12^+$ &$11^-$ & 2.424 &$14^+$ &$11^-$ & 2.566 &
$13^-$ &$12^+$ & 1.689 \\
$ 5^-$ &$ 3^-$ & 1.246 &$13^-$ &$12^+$ & 2.515 &$15^-$ &$12^+$ & 2.632 &
$14^+$ &$13^-$ & 1.720 \\
$ 7^-$ &$ 5^-$ & 1.413 &$14^+$ &$13^-$ & 2.602 &$16^+$ &$13^-$ & 2.694 &
$15^-$ &$14^+$ & 1.750 \\
$ 9^-$ &$ 7^-$ & 1.547 &$15^-$ &$14^+$ & 2.685 &$17^-$ &$14^+$ & 2.752 &
$16^+$ &$15^-$ & 1.777 \\
$11^-$ &$ 9^-$ & 1.658 &$16^+$ &$15^-$ & 2.764 &$18^+$ &$15^-$ & 2.806 &
$17^-$ &$16^+$ & 1.804 \\
$13^-$ &$11^-$ & 1.752 &$17^-$ &$16^+$ & 2.840 &$19^-$ &$16^+$ & 2.856 &
$18^+$ &$17^-$ & 1.829 \\
$15^-$ &$13^-$ & 1.832 &$18^+$ &$17^-$ & 2.913 &$20^+$ &$17^-$ & 2.904 & 
$19^-$ &$18^+$ & 1.852 \\
$17^-$ &$15^-$ & 1.902 &$19^-$ &$18^+$ & 2.983 &       &       &       &
$20^+$ &$19^-$ & 1.875 \\
$19^-$ &$17^-$ & 1.964 &$20^+$ &$19^-$ & 3.050 &       &       &       &
       &       &       \\
\hline
\end{tabular}
\end{table}


\begin{figure}[ht]
\center{\includegraphics[height=90mm]{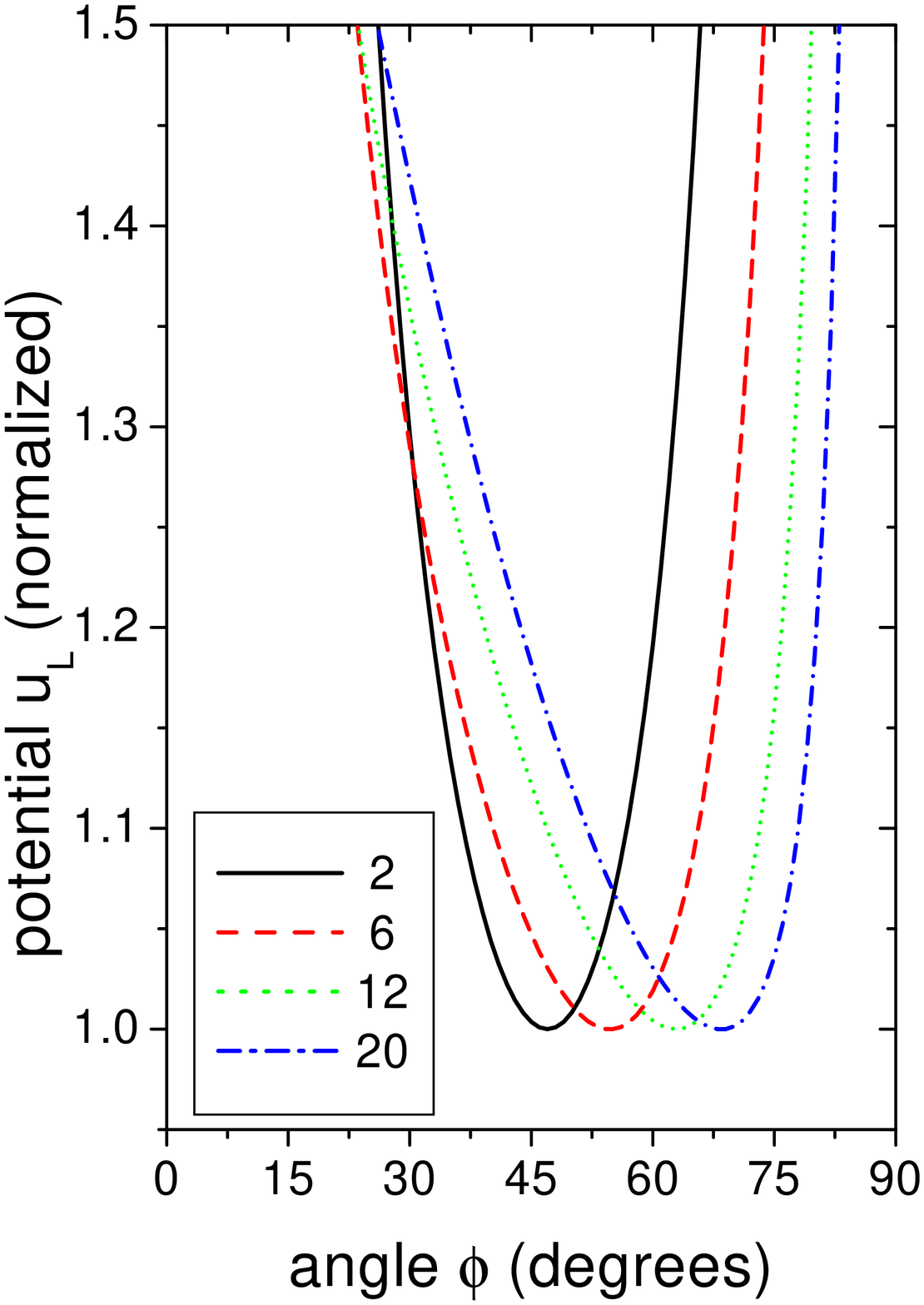}} 
\caption{
Potential $u_L$ [Eq. (\ref{eq:e8})] for different values of the angular 
momentum $L$, normalized for each $L$ to its minimum value. See section 2
for further discussion. }
\end{figure}


\begin{figure}[ht]
\rotatebox{270}{\includegraphics[height=80mm]{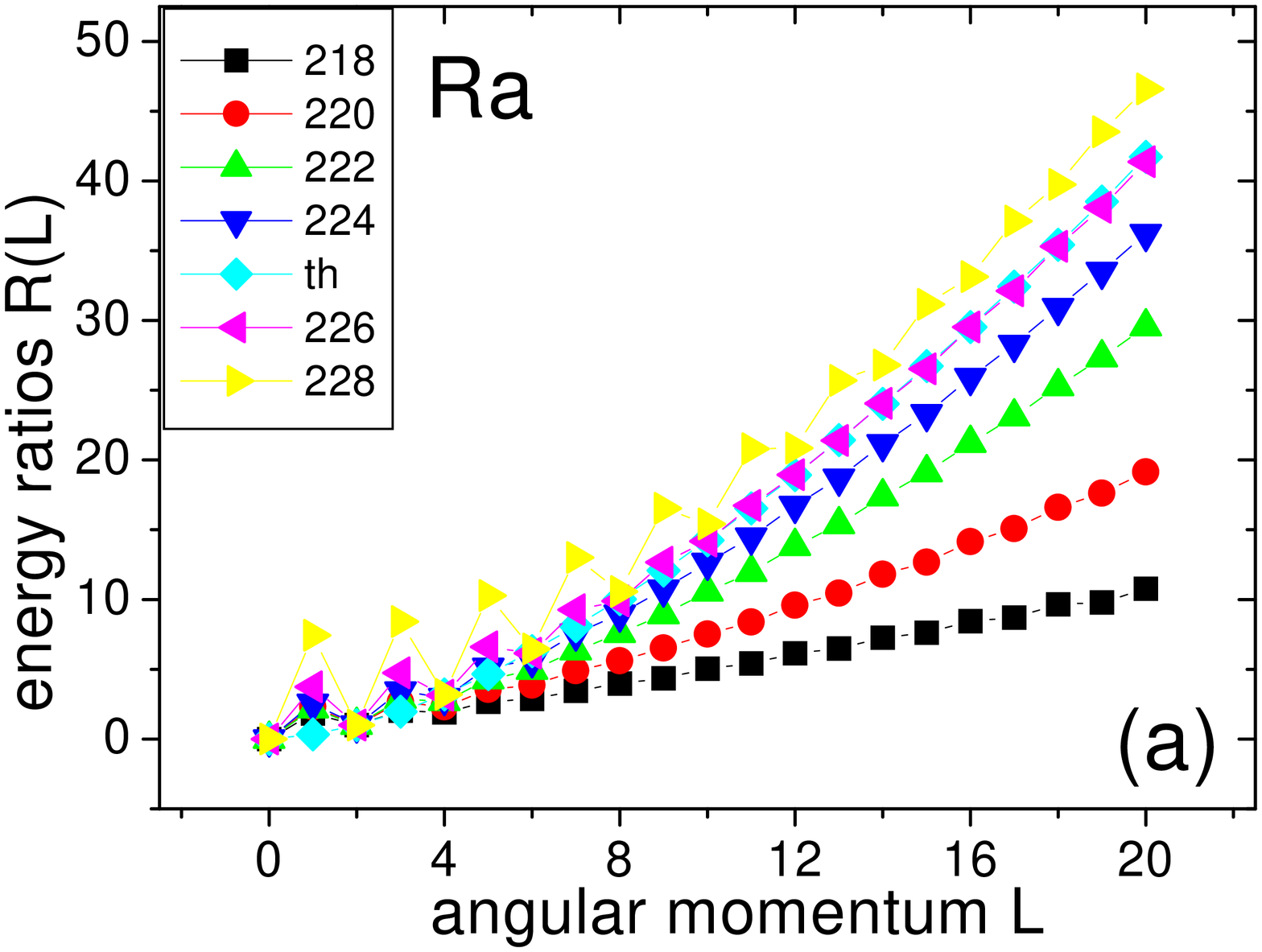}} 
\rotatebox{270}{\includegraphics[height=80mm]{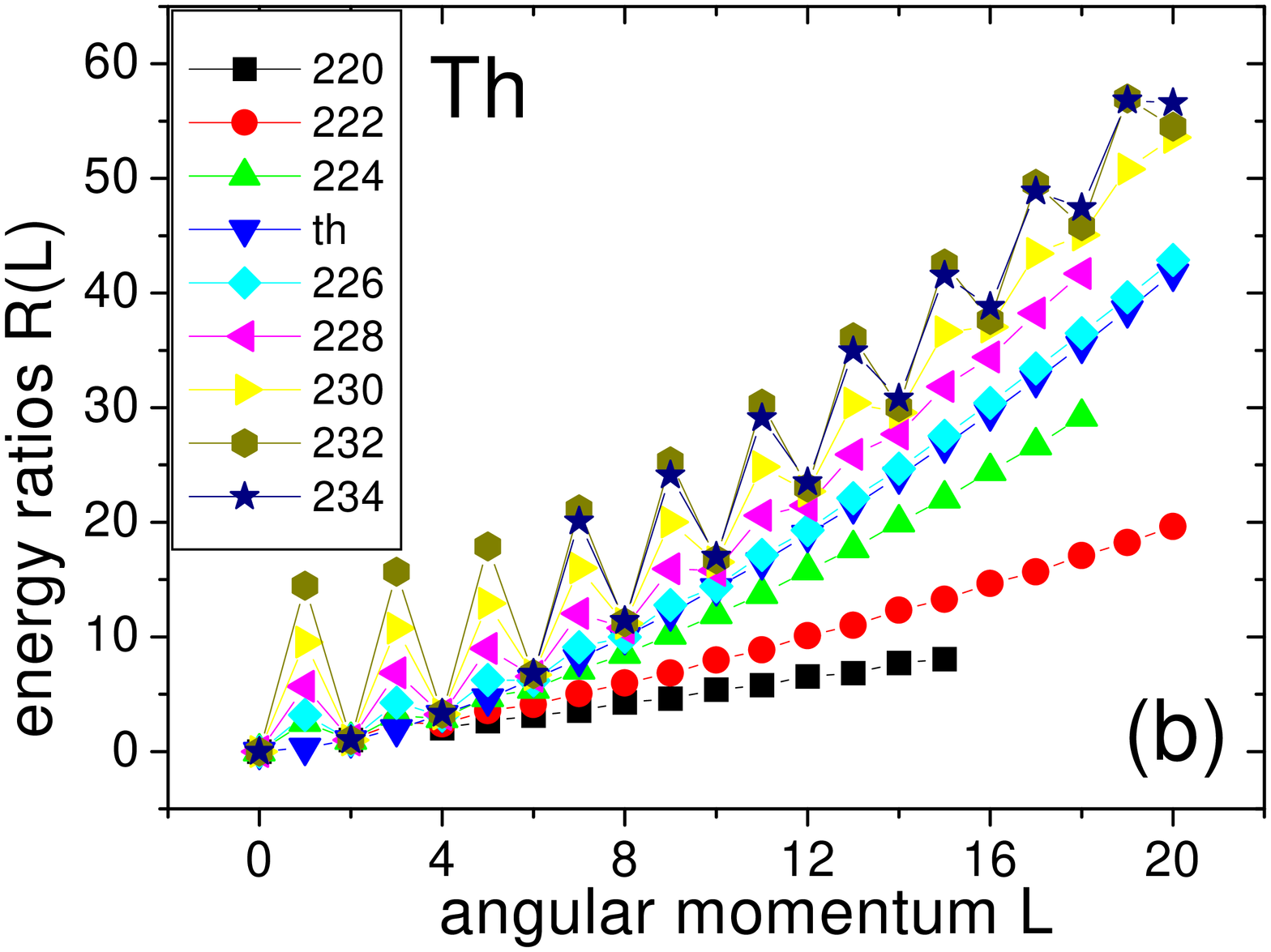}} 
\caption{
(a) Experimental energy ratios $R(L)=E(L)/E(2_1^+)$ for 
$^{218}$Ra \cite{NDS,Schulz}, $^{220}$Ra \cite{NDS}, and 
$^{222-228}$Ra \cite{Cocks}, compared to theoretical predictions.
(b) Same for $^{220-228}$Th \cite{NDS}, $^{230}$Th \cite{Cocks}, $^{232}$Th 
\cite{Cocks,NDS}, and $^{234}$Th \cite{Cocks}.}
\end{figure}  


\begin{figure}[ht]
\rotatebox{270}{\includegraphics[height=80mm]{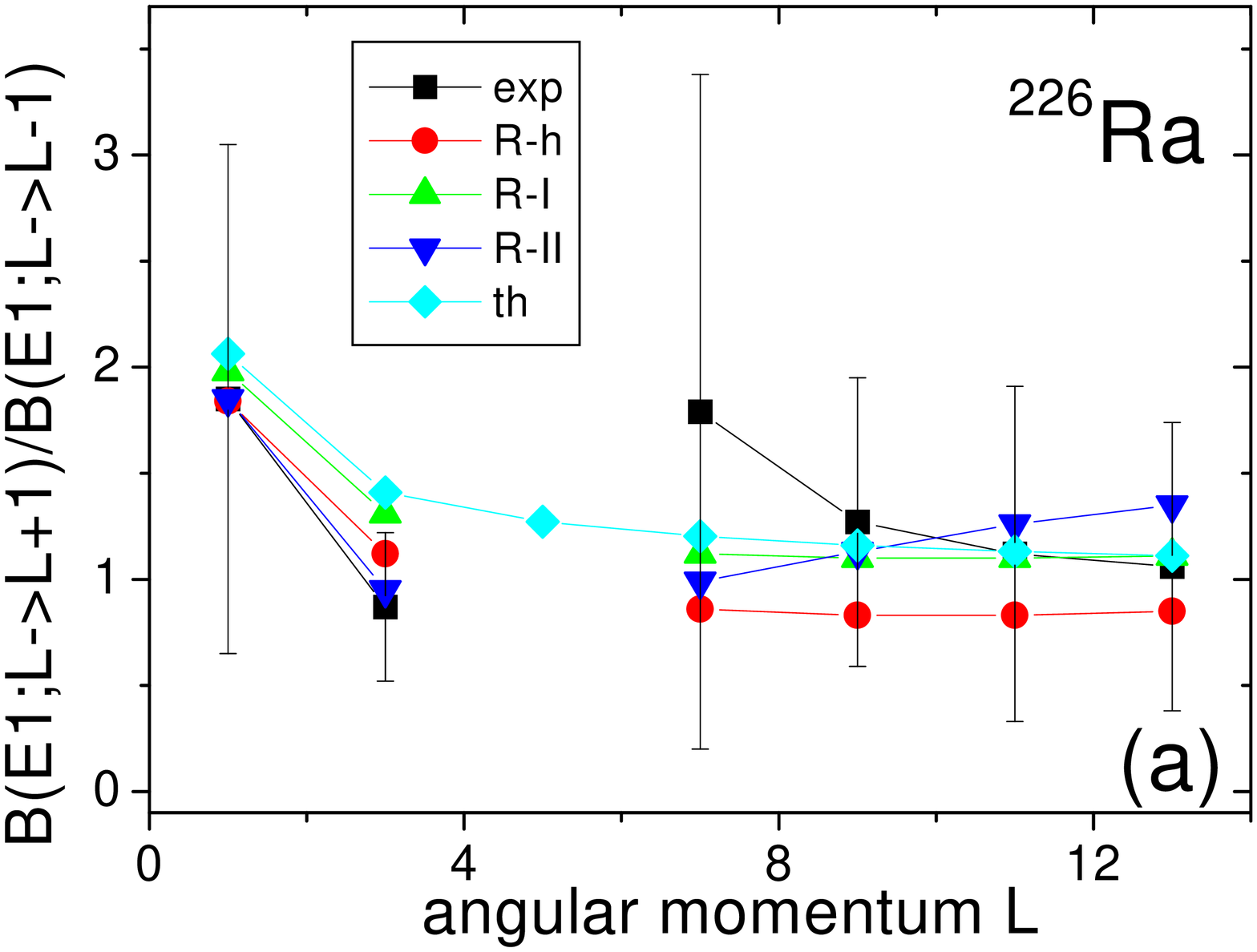}} 
\rotatebox{270}{\includegraphics[height=80mm]{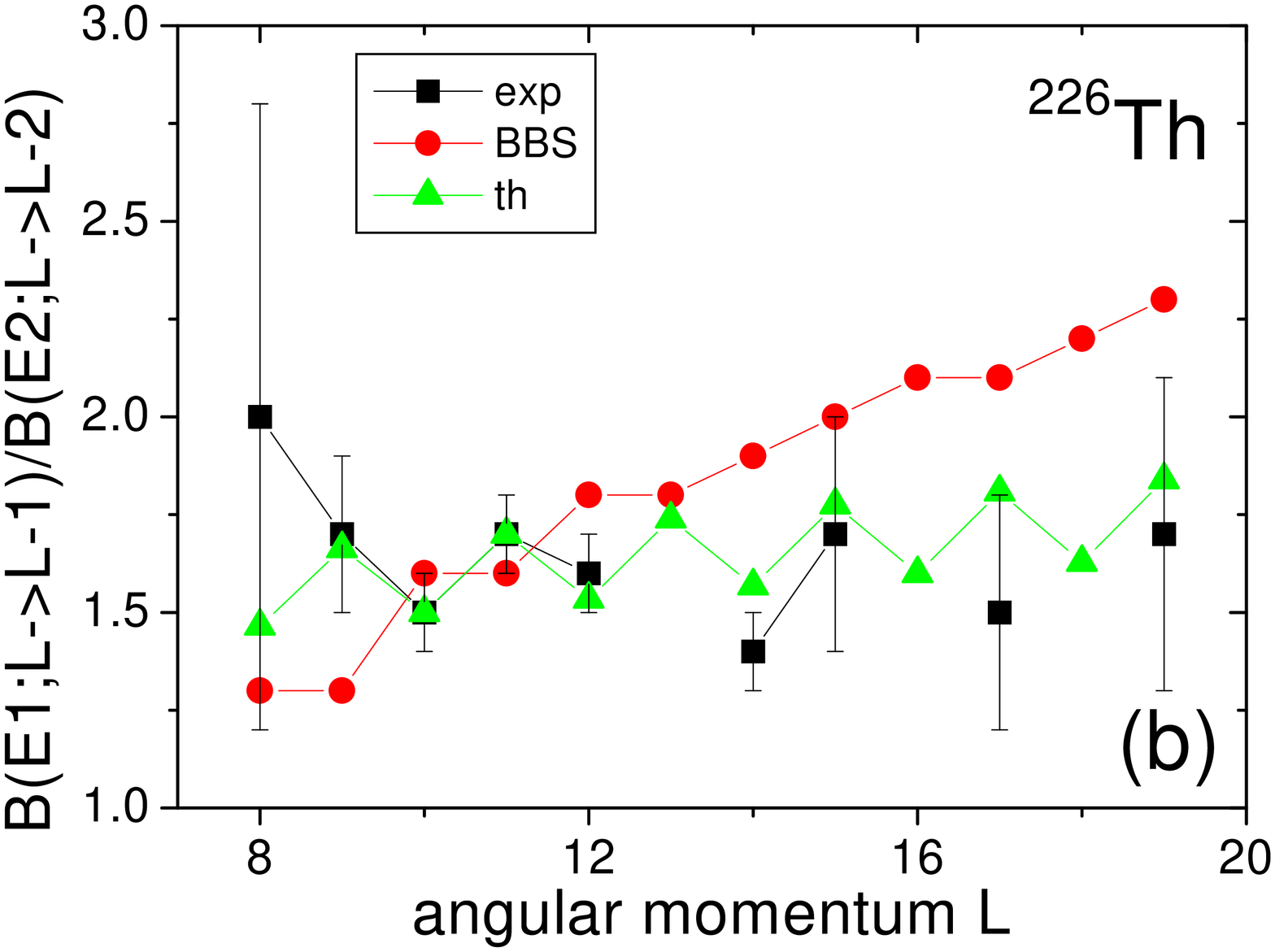}} 
\caption{
(a) Experimental $B(E1; L\to L+1)$ / $B(E1; L\to L-1)$ ratios \cite{Woll}
of $B(E1)$ values originating from the same level of $^{226}$Ra, compared 
to three different theoretical predictions from Ref. \cite{Raduta2} (labeled 
as R-h, R-I, R-II), as well as to theoretical predictions of the present 
model. See section 3 for further discussion.  
(b) Experimental $B(E1; L \to L-1)$ / $B(E2; L \to L-2)$ ratios
(multiplied by $10^5$) \cite {Bizzeti} of $B(E1)$ and $B(E2)$
values originating from the same level of $^{226}$Th, compared to theoretical 
predictions 
by Bizzeti and Bizzeti-Sona \cite{Bizzeti} (labeled as BBS), as well as to 
predictions of the present model. The ratios corresponding to $L=10$ and 11 
have been used for normalization, as in Ref. \cite{Bizzeti}. }
\end{figure}  


\begin{thebibliography}{99}

\bibitem{IacE5}
F. Iachello, Phys. Rev. Lett. 85 (2000) 3580.

\bibitem{IacX5}
F. Iachello, Phys. Rev. Lett. 87 (2001) 052502. 

\bibitem{CZE5}
R. F. Casten and N. V. Zamfir, Phys. Rev. Lett. 85 (2000) 3584. 

\bibitem{ClarkE5}
R. M. Clark, et al., Phys. Rev. C 69 (2004) 064322. 

\bibitem{CZX5}
R. F. Casten and N. V. Zamfir, Phys. Rev. Lett. 87 (2001) 052503. 

\bibitem{ClarkX5}
R. M. Clark, et al., Phys. Rev. C 68 (2003) 037301. 

\bibitem{Bohr}
A. Bohr, Mat. Fys. Medd. K. Dan. Vidensk. Selsk. 26 (1952) no. 14. 

\bibitem{RoweI}
D. J. Rowe, Nucl. Phys. A {\bf 745}, 47 (2004). 

\bibitem{RoweII}
P. S. Turner and D. J. Rowe, Nucl. Phys. A {\bf 756}, 333 (2005). 

\bibitem{RoweIII}
G. Rosensteel and D. J. Rowe, Nucl. Phys. A {\bf 759}, 92 (2005). 

\bibitem{BN}
P. A. Butler and W. Nazarewicz, Rev. Mod. Phys. 68 (1996) 349. 

\bibitem{VMI}
M. A. J. Mariscotti, G. Scharff-Goldhaber, and B. Buck, Phys. Rev. 178 
(1969) 1864. 

\bibitem{Cocks} 
J. F. C. Cocks, et al., Nucl. Phys. A  645 (1999) 61. 

\bibitem{NDS}
Nuclear Data Sheets, as of December 2004. 

\bibitem{Bizzeti}
P. G. Bizzeti and A. M. Bizzeti-Sona, Phys. Rev. C  70 (2004) 064319.

\bibitem{AQOA}
D. Bonatsos, D. Lenis, N. Minkov, D. Petrellis, and P. Yotov, Phys. Rev. C 
71 (2005) 064309. 

\bibitem{Dzy} 
A. Ya. Dzyublik and V. Yu. Denisov, Yad. Fiz. 56 (1993) 30 [Phys. 
At. Nucl. 56 (1993) 303]. 

\bibitem{Den}
V. Yu. Denisov and A. Ya. Dzyublik, Nucl. Phys. A 589 (1995) 17.

\bibitem{BM}
A. Bohr and B. R. Mottelson, Nuclear Structure, Vol. II, Benjamin, New York,
1975. 

\bibitem{Fortun}
L. Fortunato, Phys. Rev. C 70 (2004) 011302(R). 

\bibitem{Fiedel}
H. Fiedelday amd S. A. Sofianos, Z. Phys. A  311 (1983) 339. 

\bibitem{Scheid}
W. Greiner, J. Y. Park, and W. Scheid, Nuclear Molecules, World Scientific,
Singapore, 1995. 

\bibitem{Schulz} 
N. Schulz, et al., Phys. Rev. Lett. 63 (1989) 2645. 

\bibitem{Jolos}
R. V. Jolos and P. von Brentano, Phys. Rev. C 49 (1994) R2301. 

\bibitem{Woll}
H. J. Wollersheim, et al., Nucl. Phys. A 556 (1993) 261. 

\bibitem{Raduta} 
A. A. Raduta, Recent Res. Devel. Nuclear Phys.  1 (2004) 1. 

\bibitem{Raduta2}
A. A. Raduta, D. Ionescu, I. I. Ursu, and A. Faessler, Nucl. Phys. A 
 720 (2003) 43.

\bibitem{X5}
D. Bonatsos, D. Lenis, N. Minkov, P. P. Raychev and P. A. Terziev, 
Phys. Rev. C  69 (2004) 014302.  

\end{thebibliography}
\end{document}